\documentclass[amsmath,amssymb,aps,prl,twocolumn,groupedaddress]{revtex4-2}

\usepackage{graphicx}
\usepackage{dcolumn}
\usepackage{bm}

\begin{document}


\title{Extraordinary Quality Factors in Dual-Band Polarization-Insensitive QuasiBound States in the Continuum}



\author{Maryam Ghahremani}
\email[]{ghahremani.maryam@ut.ac.ir}
\affiliation{Photonics Research Laboratory, Center of Excellence on Applied Electromagnetic Systems, School of Electrical and Computer Engineering, College of Engineering, University of Tehran, Tehran, Iran}

\author{Carlos J. Zapata-Rodr\'{\i}guez}
\email[]{carlos.zapata@uv.es}
\affiliation{Department of Optics and Optometry and Vision Sciences, Faculty of Physics, University of Valencia, Dr. Moliner 50, Burjassot 46100, Spain}


\date{\today}

\begin{abstract}
	
In this study, we investigate a novel ``dimerized'' dielectric metasurface featuring dual-mode resonances governed by symmetry-protected bound states in the continuum (BICs). 
The metasurface design offers advantages such as insensitivity to incident light polarization and exceptionally high quality factors exceeding 10$^5$ for low and moderate structural deviations from the monoatomic array. 
By introducing mastered perturbations in the metasurface via the Brillouin zone folding method, without reducing symmetry, we explore the behavior of symmetry-protected BIC states and their polarization-independent responses. 
Through numerical simulations and analysis, we demonstrate the superiority of $Q$ factors for specific BIC-based resonances, leading to precise control over interaction behaviors and light engineering in both near and far fields. 
Our findings contribute to the understanding of BIC resonance interactions and offer insights into the design of high-performance sensing applications and meta-devices with enhanced functionalities.
 
\end{abstract}


\maketitle

\section{}

\section{Introduction}

Metasurfaces consisting of two-dimensional arrays of meta-atoms with a thickness of less than the wavelength are the focus of current research activities for their exceptional ability to effectively manipulate and control light-matter interactions in an ultra-narrow spectral window~\cite{Sautter15,Genevet17,Hsiao17,Li18}. 
This characteristic stems from a resonant  effect, which are mostly controlled by the geometry of the unit cells.
Thus, metasurfaces are capable of generating a variety of unique electromagnetic responses. 
These include anapole mode~\cite{Algorri18,Baryshnikova19,Ghahremani21}, Fano resonance~\cite{Cong15,Campione16,Limonov17}, and electromagnetically-induced transparency~\cite{Yang14,Yahiaoui18,Diao19}. 
The unique capabilities of metasurfaces have led to planar optical meta-components with a wide range of applications across both microwave and optical frequencies in the fields of imaging and lensing~\cite{Hashemi16,Khorasaninejad17,Shanei17,Shrestha18}, holography~\cite{Ni13,Zheng15,Wen15}, structural color printing~\cite{Cheng15b,Sun17,Yang20}, energy harvesting~\cite{Zhang17,Ghaderi18,Li20}, and perfect absorbers~\cite{Akselrod15,Raad20,Karimi22}, among others. 
Importantly, the recent development of all-dielectric metasurfaces especially benefit from highly efficient transmission applications~\cite{Jahani16,Kamali18,Koshelev21}. 
Unlike their plasmonic counterparts, all-dielectric metasurfaces, in addition to their compatibility with standard complementary metal-oxide semiconductor (CMOS) foundries, do not suffer from substantial material losses and can thus support Mie-type electric and magnetic resonances with extremely-high quality ($Q$) factors. 

In recent years, a new design paradigm to achieve high-$Q$ resonances has been recognized which is based on the physics of bound states in the continuum (BICs). 
The existence of such peculiar resonant states was initially demonstrated by Neumann and Wigner~\cite{Von93} in the context of quantum mechanics; 
however, many photonic structures, such as dielectric photonic crystals (PhCs), have been widely discussed for optical BICs in the last decade~\cite{Li19,Meng22,Mohamed22,Xiao22,Manez24}.
In PhC slabs, an optical BIC state (also referred as trapped modes) corresponds to an eigenmode that is ideally guided, i.e., completely decoupled from the outgoing waves, despite the fact that it occurs above the light line in the dispersion diagram $\omega = \omega (\vec{k})$ with a resonance frequency $\omega$ and $\vec{k}$ the in-plane wave-vector~\cite{Hsu13}. 
In principle, the eigenmodes existing above the light line should be leaky, i.e., it expects them to decay as they propagate owing to radiative leakage. 
The absence of radiation, the fact that the state is bound (perfectly localized), in lossless infinite structures provides infinite $Q$-factor resonances with zero linewidth which are not observable in the far-field spectra due to the lack of radiative channels. 
Therefore, the ideal BIC states cannot be manifested in real practical cases and are rather mathematical concepts. 
In real non-ideal structures, BIC modes can be converted into a resonant condition, the so-called quasi-bound states in the continuum (Q-BIC), due to technological imperfections, surface roughness, material dissipative losses, or finite size of the device. 

The optical BIC modes are classified in several different categories, one of which is referred as the symmetry-protected BIC (SP-BIC) existed at the $\Gamma$-point of the first Brillouin zone (FBZ) in PhCs. 
The SP-BIC state emerges from the symmetry mismatch between the modal field profile and the external outgoing wave. 
In dielectric PhC slabs, an ideal SP-BIC state can be transformed into a leaky resonance through introducing a slight perturbation that breaks the metasurface unit-cell configuration symmetry. 
Using an appropriate structural symmetry reduction allows field distribution of the SP-BIC mode is perturbed and transformed to a Q-BIC mode that can couple to the symmetric profile of normally-incident plane waves propagating as the radiation channel.
Through manipulating the introduced asymmetry in the unit cell, one can effectively control and engineer radiative losses from Q-BIC in periodic structures~\cite{Koshelev18} to attain the desired high-$Q$ states. 
This mechanism of transition from SP-BICs to Q-BICs can offer a novel route to achieve ultrahigh $Q$-factor resonances with greatly enhanced electromagnetic near-fields in the metasurface plane which makes it exceptionally important for the applications in refractive index sensors~\cite{Gao23} and non-linear optics~\cite{Liu19,Koshelev19}. 

\begin{figure*}[t]
	\includegraphics[width=0.75\textwidth]{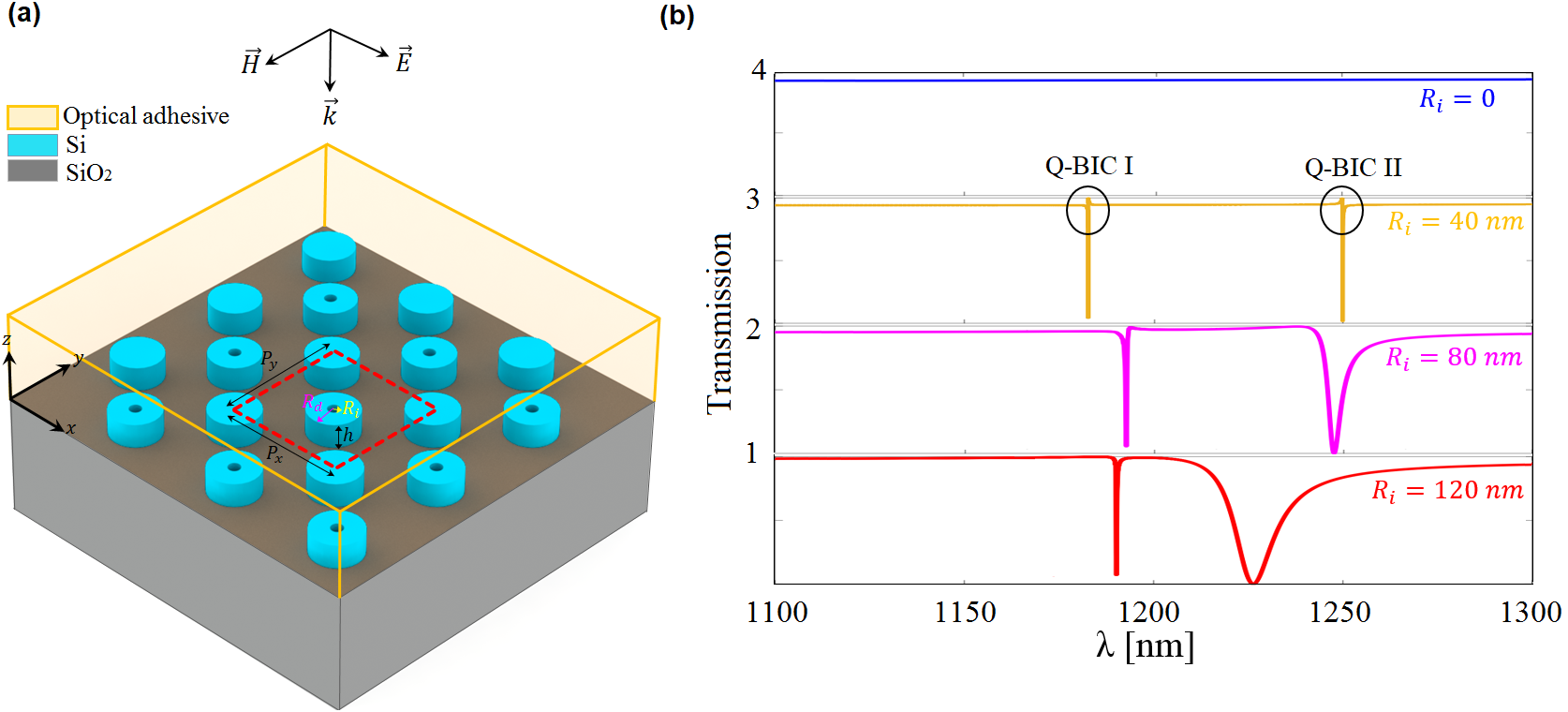}
	\caption{\label{fig:01} 
		(a) Schematics of the metasurface.
		(b) Spectral transmittance under $x$-polarized normal incidence within the region of interest for diferent radii $R_i$ of the inner hole.
	}
\end{figure*}

Resonances governed by SP-BIC states commonly exhibit sensitivity to the polarization of the incident light because the in-plane or out-of-plane asymmetry required for their excitation always damages the C$_{4v}$ rotational symmetry or mirror symmetry of the structures. 
To meet the challenge of designing metasurfaces with polarization-independent responses have gained considerable attention in recent years due to the flexible practical applications. 
In this regard, it was proposed to utilize cluster-based modifications of metasurfaces where ordinary particles are grouped together to form a unit super-cell (cluster)~\cite{Chen23}. 
To excite optical trapped modes, design of the cluster forming the two-dimensional lattice can be realized in patterns based on either the equidistantly-spaced particles with perturbations in geometry or refractive index, or using the single-size non-perturbed particles with specific displacements toward the super-cell center.
To characterize fundamental properties of such metasurfaces, group-theoretical methods solely based upon the symmetries of their unit clusters are employed. 
The particles within the super-cell can be arranged in accordance with the symmetry groups C$_{4v}$, C$_{2v}$, C$_{4}$, C$_{2}$ and C$_{s}$ studied in~\cite{Yu19,Overvig20}. 
Among them, the symmetry perturbation methods maintaining maximal symmetry group C$_{4v}$ is more encouraging. 
For most of the previously reported such super-cell-based metasurfaces, to achieve the high-$Q$-factor resonances close to 10$^6$ or more, which are quite promising particularly for refractive index sensing applications, requires setting a very small asymmetry parameter in the range of 0.005 or less. 
In this case, practical realizations of the designs are rather rare and sometimes impossible.  
As the perturbation parameter becomes large, the $Q$ factors of Q-BICs decrease significantly. 
Therefore, a long-standing question that remains unresolved is: 
how to achieve ultrahigh-$Q$ Q-BICs when the perturbation parameter varies within a large range? 


In this paper, we propose a ``dimerized'' dielectric metasurface with dual-mode resonances governed by SP-BIC states which offer two main advantages. 
First, they are insensitive to the normally-incident light with any state of polarization mapped onto the Poincaré sphere. 
Second, they represent $Q$ factors of more than 10$^6$ for the asymmetry parameter about $\alpha = 0.05$, which is at least 10 times larger than the previously reported designs. 
Such large values of the asymmetric differences avoids limitations of current fabrication process technology. 
From the study of the photonic band structure, we theoretically investigate the result of Brillouin zone folding and relocation of the BIC states when we introduce a perturbation into the metasurface structure through thoroughly perforating a concentric hole in half of the nanodisks. 
Upon this feature, the eigenmodes of the unperturbed lattice existed in the guided region zone are brought into the continuum to form Q-BIC states. 
We consider that the metasurface under study is composed of silicon (Si) nanodisk resonators ordered in a square lattice and supporting C$_{4v}$ high-symmetry features, subsequently providing extremely-high $Q$ factors. 
Our proposed metasurface with simpler fabrication process and in a fully symmetric arrangement may facilitate design and realization of ultrahigh-performance sensing applications.

\section{Device design and numerical simulations}

Figure~\ref{fig:01}(a) shows the schematics of the metasurface that is the object of our analysis.
This metasurface is based on the controlled perturbation upon a square array of silicon disks with lattice constant $P'$.
The dielectric disks have a radius $R_d$ and a height $H$, lying on a SiO$_2$ substrate and being cladded by an optical adhesive, both occupying a semi-infinite space for simplicity.
If not mentioned explicitly, we will consider $R_d = 165$~nm and $H = 160$~nm in our simulations.
Moreover, the index of refraction of Si and SiO$_2$ is here taken from Ref.~\cite{Palik99}, which are valid within the near-infrared window of our interest ranging from $1100$~nm and $1300$~nm.
The optical adhesive layer is here considered of refractive index $n_\text{ad} = 1.33$.
Next a disturbance is introduced in our primary periodic structure by etching a concentric hole of radius $R_i$ in intersperced Si disks, as depicted in Fig.~\ref{fig:01}(a).
This procedure retains the C$_{4v}$ in-plane symmetry of the original metasurface, thus enabling an optically isotropic response~\cite{Yu19,Vaity22}.
In addition, the topology of such a disturbance enables the generation of extremely-narrow resonances when illuminated by an $x$-polarized plane wave, as shown in Fig.~\ref{fig:01}(b) for $R_i > 0$.
Our purpose is focused on illustrating how the strategy here followed drives a strong control over the $Q$ factors of specific Q-BIC-based resonances found in the transmission response.
On this subject, in addition, it remains challenging to demonstrate high-$Q$ Q-BICs in experiments because ultra-small structural perturbations are demanding to fabricate with state-of-the-art techniques, and because inevitable fabrication imperfections introduce scattering losses.~\cite{Lee21,Gao22,Zhang23}.

In Fig.~\ref{fig:01}(b) we show the transmittance of the metasurface when injecting a $x$-polarized plane wave under normal incidence conditions within the spectral region comprised between $1100$~nm and $1300$~nm.
The numerical simulations were performed using the commercial software COMSOL Multiphysics, which is a solver of the Maxwell's equations based on the finite element method~\cite{COMSOL-FEM}.
In virtue of the C$_{4v}$ symmetry of our structure, the results are replicated by $y$-polarized plane waves.
The unperturbed metasurface characterized by $R_i = 0$~nm demonstrates a flat response revealing a near-unity transmittance in the spectrum of interest.
However, a disturbance in the original periodic distribution of the arrays introduces the signature of a pair of Q-BICs, where transmittance drastically drops within a very narrow spectral band.
This fact is a manifestation that symmetry-protected BICs cannot be observed through prescribed electromagnetic signals because of its nonradiative intrinsic features with vanishing resonance linewidths.
In the case of $R_i = 40$~nm, both Q-BICs are revealed by corresponding minima in transmission, and therefore maxima in reflection, centered at $1181$~nm and $1250$~nm;
from here on they will be coined as Q-BIC I and Q-BIC II, respectively.
When increasing the radius of the centered hole by $R_i = 80$~nm, Q-BIC I is redshifted however minimizing the degradation of the sharp peak-and-trough (Fano) profile associated with this resonance.
On the contrary, Q-BIC II keeps the central frequency of the sharp resonance however increasing the curve width notably.
Moreover, when the inner radius $R_i$ increases up to 120~nm, Q-BIC I is still resilient upon a resonance degradation, in opossition to the case of Q-BIC II.
This unique characteristic of Q-BIC I, which in addition is polarization independent, constitutes the main result of our study.

\begin{figure*}[t]
	\includegraphics[width=0.75\textwidth]{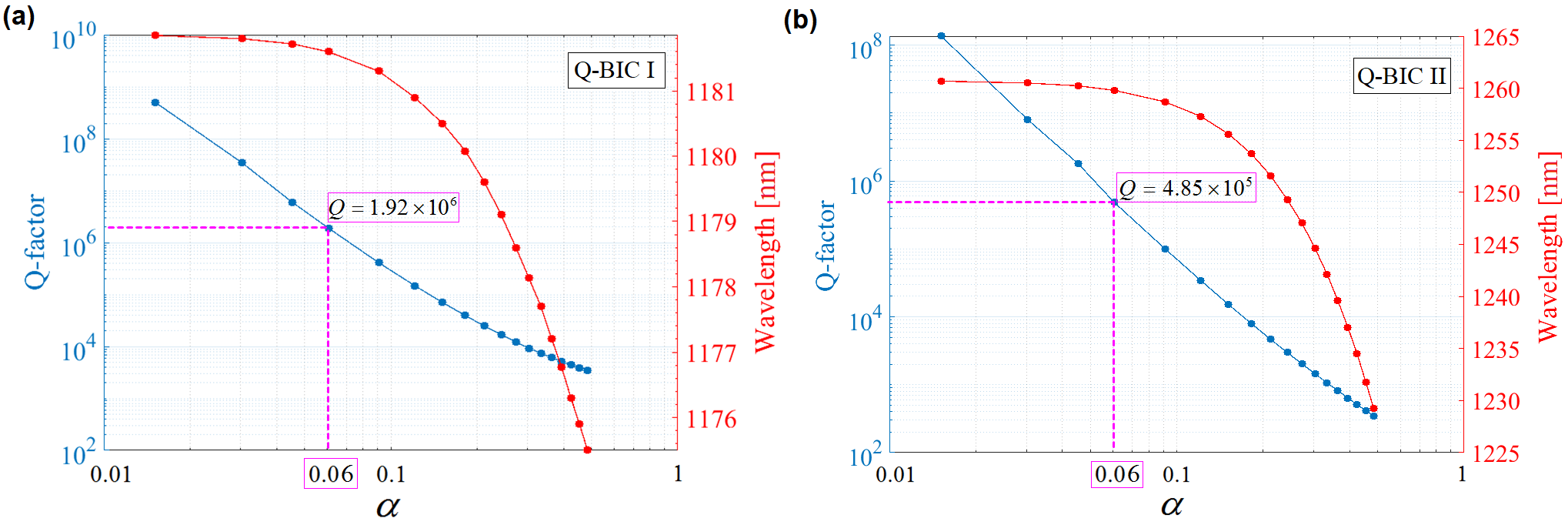}
	\caption{\label{fig:02} 
		Dependence of the quality factor and central wavelength on the asymmetry parameter $\alpha$ for: (a) Q-BIC I and (b) Q-BIC II.
	}
\end{figure*}

The $Q$ factor can be estimated through the transmission spectrum.
More especifically, the Fano model well fits the spectral features of the transmittance as~\cite{Limonov17,Cai21,Fan22},
\begin{equation} \label{eq:01}
	T (\omega)
	=
	T_0
	+
	A_0
	\frac{[q + 2 (\omega - \omega_0)/\gamma]^2}{1 + [2 (\omega - \omega_0)/\gamma]^2} ,
\end{equation}
where $\omega_0$ is the resonant frequency, $\gamma$ is the resonance linewidth, $T_0$ is the transmission offset, $A_0$ is the continuum-discrete coupling constant, and $q$ is the Breit-Wigner-Fano parameter determining asymmetry of the resonance profile. 
By estimating the values $\omega_0$ and $\gamma$ through a curve fitting, the $Q$ factor is determined using the expression~\cite{Wang23b}
\begin{equation}
	Q
	=
	\frac{\omega_0}{2 \gamma} .
\end{equation}
Especifically for Q-BIC I and Q-BIC II, their corresponding $Q$ factors reach $1.7 \times 10^4$ and $3.0 \times 10^3$ for $R_i = 40$~nm, respectively, however droping to $3.4 \times 10^3$ and $3.4 \times 10^2$ when  $R_i = 80$~nm.
In fact, the resulting peak positions and linewidths correspond exactly to the
real and imaginary parts of the eigenmode frequencies~\cite{Koshelev18}, as we will corroborate down this lines. 

A convenient procedure at this point lies on representing the $Q$-factor variation in terms of the so-called asymmetry parameter $\alpha$, which can be defined for our structure by
\begin{equation}
	\alpha
	=
	\frac{R_i}{R_d} .
\end{equation}
Note that $\alpha = 0$ is the minimum value of the asymmetry parameter that corresponds to the undisturbed lattice, a case providing a $Q$ factor ideally tending to infinity.
Otherwise $\alpha \le 1$ is always satisfied.
In particular, $\alpha \approx 6 \times 10^{-2}$ if $R_i = 10$~nm, which is a value highlighted in Fig.~\ref{fig:02}, the latter illustrating the variation of the $Q$ factor in terms of $\alpha$ for Q-BIC I and Q-BIC II.
Note that an alternate definition $\alpha = {\Delta S} / {S}$ can be found elsewhere~\cite{Koshelev18,Wang20b}, where $\Delta S = \pi R_i^2$ and $S = \pi R_d^2$ are the areas of the centered hole and the nanodisk, respectively.
This fact will be taken into account in our comparative (see Table~\ref{tab:01}).
Our results endorse the well-known inverse square dependence of the $Q$ factor upon the asymmetry parameter $\alpha$, namely~\cite{Koshelev18}
\begin{equation} \label{eq:05}
	Q
	=
	\frac{Q_0}{\alpha^2} ,
\end{equation}
where $Q_0$ is a constant determined by the metasurface design.
Equation~\eqref{eq:05} is an accurate approach for low values of the asymmetry parameter, typically $\alpha < 0.1$.
An extraordinary $Q$ factor is then related to ultrahigh values of the constant $Q_0$, which in certainly governed by the symmetry of the array up to a certain degree.
Provided $R_i = 10$~nm, leading to a value of $\alpha = 0.06$, the $Q$ factor of Q-BIC I reaches $1.92 \times 10^6$, and Q-BIC II reduces to $Q = 4.85 \times 10^5$.
Moreover, the $Q$ factor of Q-BIC I is always higher that Q-BIC II by at least one order of magnitude for values of $\alpha < 0.9$, demonstrating a high degree of resiliance upon the introduction of the specific disturbance in the photonic structure here analyzed.
Although the $Q$ factor for the Q-BIC II resonance is lower than that of Q-BIC I, it is still considerably high for a polarization independent design.
This can be verified by comparing the $Q$ factor of analogous BIC-based proposals, as shown in Table~\ref{tab:01}.
The $Q$-factor decreases much more slowly for Q-BIC I and is maintained above $10^4$ because, as will be shown below, its eigenfield profile is weakly perturbed by the hole shape.

\begin{table*}[h]
	\caption{\label{tab:01}%
		Comparison of $Q$-factor between previous works based on multiple metasurfaces (MS) and the present study.
	}
	\begin{ruledtabular}
		\begin{tabular}{lcccc}
			\textrm{Resonant Structure} &
			\multicolumn{1}{c}{\textrm{$\alpha$}} &
			\textrm{$Q$-Factor}\footnote{Calculated} &
			\textrm{$\lambda$ (nm)} &
			\textrm{Ref.}
			\\
			\colrule
			\textrm{Dual Q-BICs on a rhomboidal MS} & 0.01 & \textrm{$10^3$} & 1055 & \cite{Qin24}
			\\
			\textrm{Permittivity-asymmetric all-dielectric MS} & 0.01 & \textrm{$5 \times 10^3$} & 1145 & \cite{Yu22}
			\\
			\textrm{Permittivity-asymmetric membrane MS} & 0.01 & \textrm{$10^4$} & 1550 & \cite{Zhou24}
			\\
			\textrm{Si nanodisk tetramer MS} & 0.02 & \textrm{$3 \times 10^4$} & 1555 & \cite{Xiao22b}
			\\
			\textrm{Amorphous-silicon (a-Si) cuboids tetramer} & 0.01 & \textrm{$10^5$} & 1265 & \cite{Cai21}
			\\
			\textrm{Dielectric cubic tetramer MS} & 0.02 & \textrm{$10^5$} & 1210 & \cite{Feng23}
			\\
			\textrm{Tetrameric silicon nanoboxes} & 0.02 & \textrm{$3 \times 10^5$} & 1300 & \cite{Gao23}
			\\
			\textrm{Our work} & 0.06 & \textrm{$2 \times 10^5$} & 1180 & 
		\end{tabular}
	\end{ruledtabular}
\end{table*}

\begin{figure*}[tb]
	\includegraphics[width=0.85\textwidth]{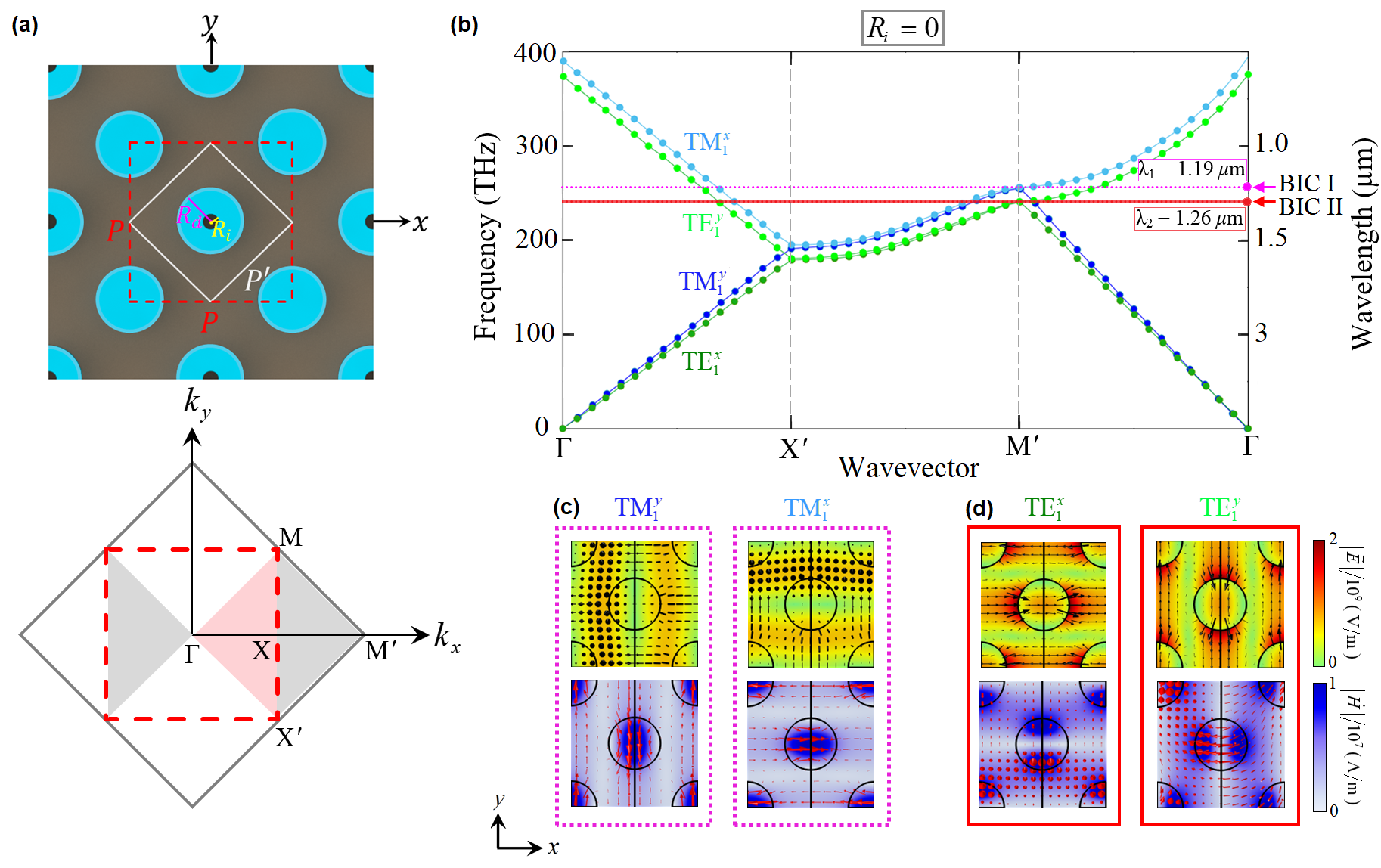}
	\caption{\label{fig:03} 
		(a) Top: First Brillouin zone in the $k_x$-$k_y$ plane of the undisturbed and dimerized square lattice, whose area contours are given by a purple dashed line and a solid green line, respectively.
		Bottom: Brillouin zones of the metasurfaces.
		(b) Band diagram of TE-like and TM-like polarized Bloch modes for the full-disk structure ($R_i = 0$~nm).
		(c)-(d) Electromagnetic fields for selected Bloch frequencies at $M'$.
	}
\end{figure*}

\section{Analysis and discussion}

\begin{figure}[tb]
	\includegraphics[width=0.45\textwidth]{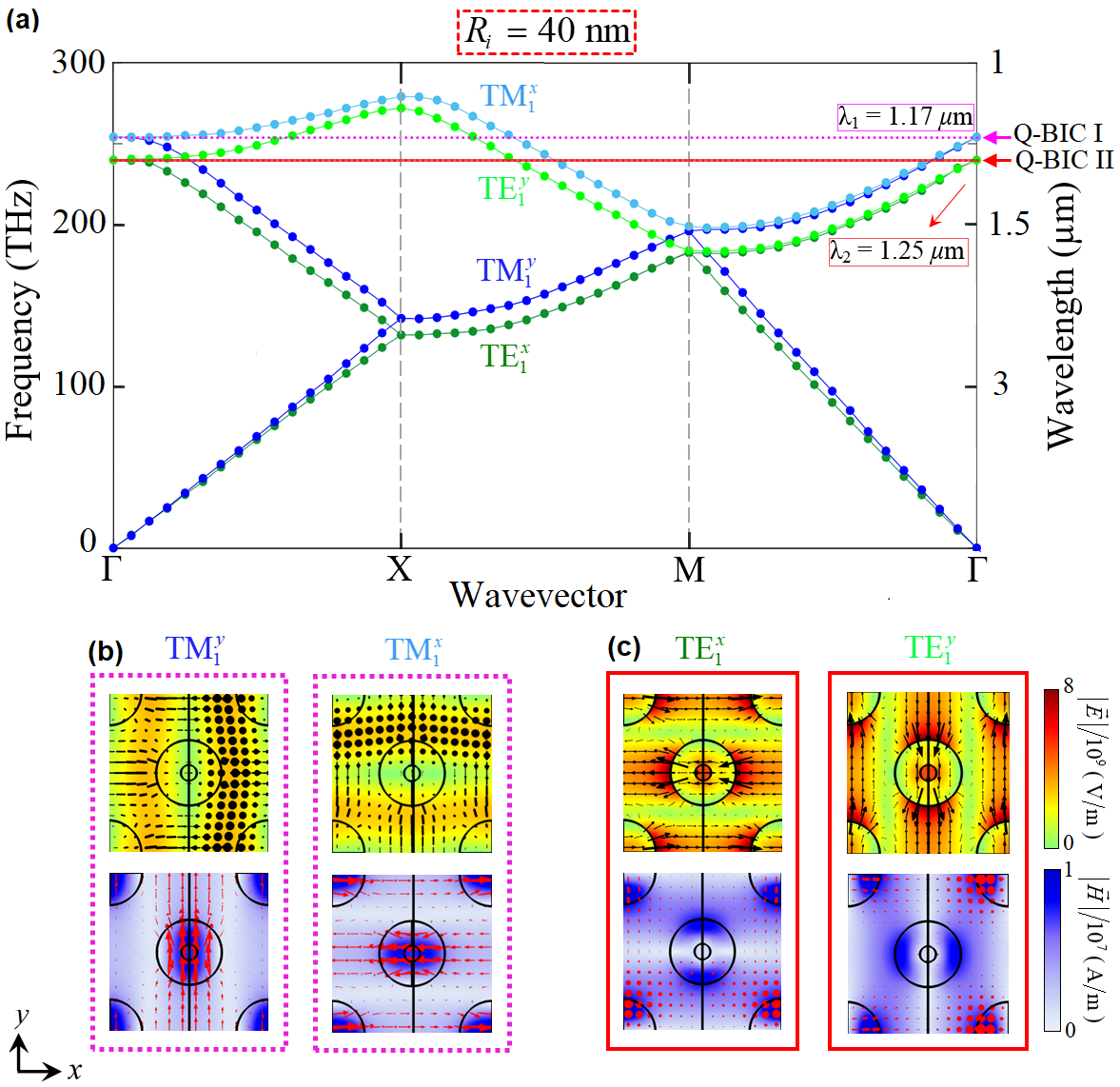}
	\caption{\label{fig:04} 
		(a) Band structure for polarized TE and TM Bloch modes for the dimerized array of meta-atoms for $R_i = 40$~nm.
		(b) and (c) Electromagnetic fields for selected Bloch frequencies.
	}
\end{figure}

\begin{figure*}[tb]
	\includegraphics[width=0.95\textwidth]{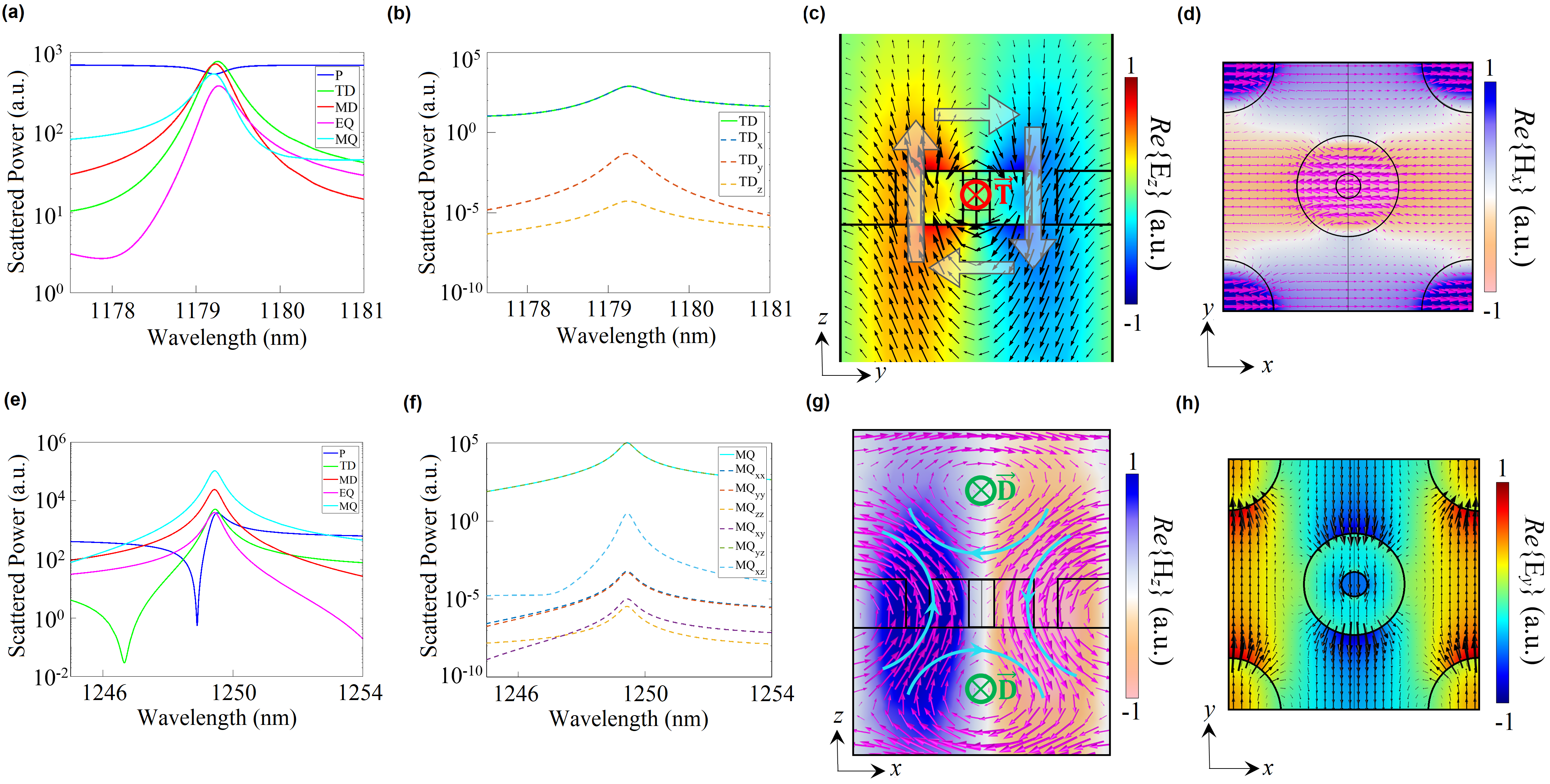}
	\caption{\label{fig:05} 
		Multipolar decomposition for the (a-b) Q-BIC I and the (e-f) Q-BIC II for the dimerized nanostructure at $R_i = 40$~nm.
		Detailed analysis of toroidal dipole components (for Q-BIC I) and magnetic quadrupole components (for Q-BIC II) are given in (b) and (f), respectively.
		Resonant electromagnetic fields around the centered holey disk of Q-BIC I (c-d) and Q-BIC II (g-h) with a natural wavelength of $1179$~nm and $1249$~nm, respectively.
	}
\end{figure*}

\begin{figure*}[t]
	\includegraphics[width=0.65\textwidth]{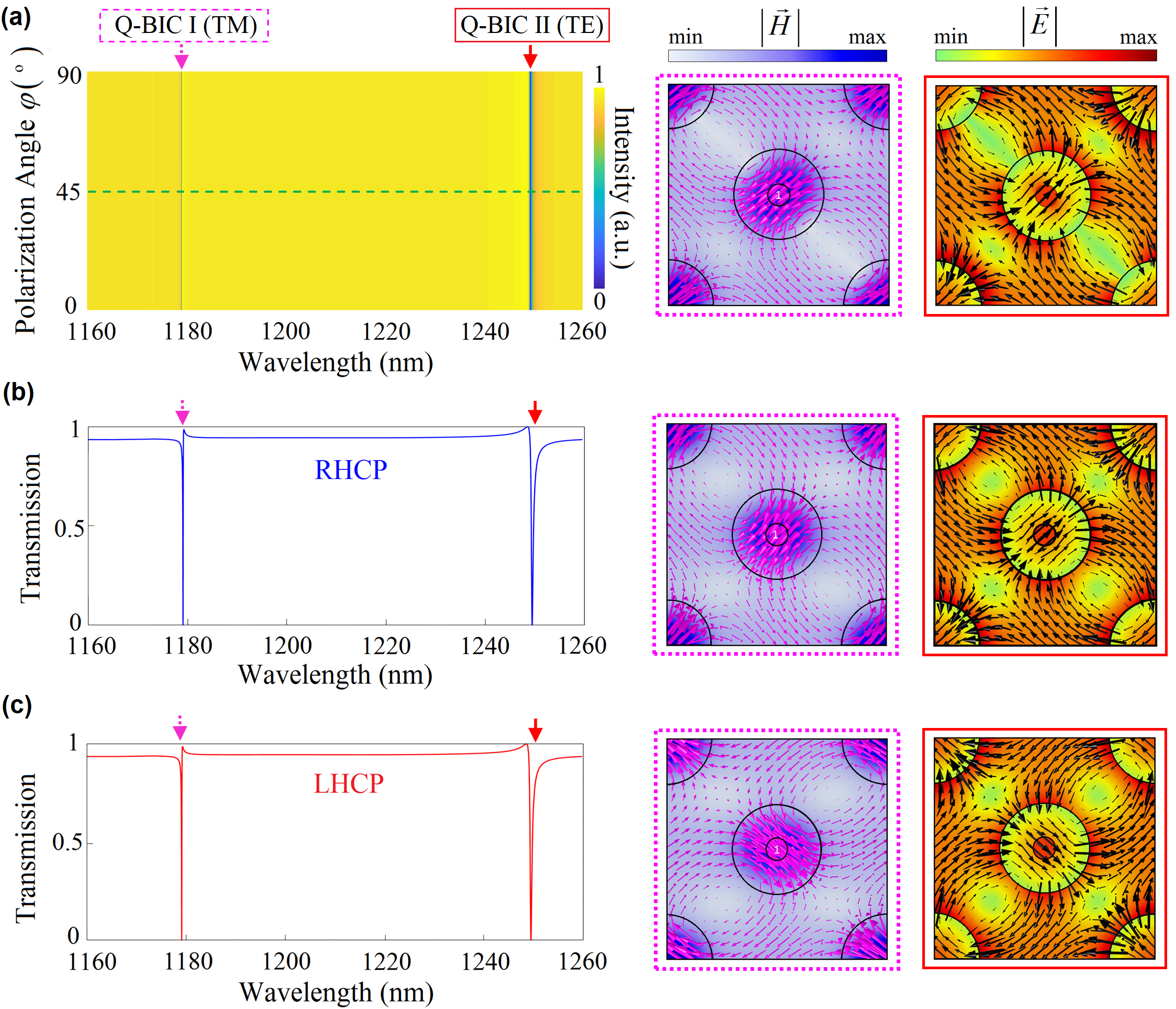}
	\caption{\label{fig:06} 
		(a) Metasurface response to different linear polarization states at $R_i = 40$~nm, showing in-plane magnetic and electric field distributions for TM-like and TE-like modes at an incident angle of $\varphi = 45^\circ$ degrees.
		Transmission spectrum and field profiles of the metasurface under right-handed (RH) (b) and left-handed (LH) (c) circularly polarized (CP) light.
	}
\end{figure*}

The disturbance executed upon the ``monoatomic'' dielectric metasurface reveals a process of dimerization in the periodic structure since the unit cell (or super-cell) now includes an unperturbed disk and a holey disk.
As a result, the lattice is reconfigured with a larger period $P > P'$ but still conserving its square lattice symmetry, as depicted in Fig.~\ref{fig:03}(a).
In our calculations, we set a holey disk at the center of the super-cell, and neighboring full disks evenly contribute by one quart each one as they are located at the super-cell corners.
Inherently, such spatial distribution manifests the dimerization of the super-cell and, in addition, explicitly retains the original C$_{4v}$ symmetry of the photonic lattice.
Note that an alternate super-cell taking the holey disks at the corners and a full Si particle set at the center will also provide the desired results.
Figure~\ref{fig:03}(a) also shows the FBZ within the $k_x$-$k_y$ Fourier plane of the square array of undisturbed Si disks ($\alpha = 0$) with boundary Bloch frequencies located at points $\Gamma$, $X'$ and $M'$.
Held within such a spectral window, the FBZ of the dimerized metasurfaces is in addition represented by the characteristic points $\Gamma$, $X$ and $M$.
Note that the spatial frequency set at $M'$ is outside the FBZ of the dimerized arrangement and, through the procedure of the Brillouin zone folding, is here translated to the point $\Gamma$.
This fact is also revealed by the inversion symmetry of the energy dispersion surfaces within the FBZ, however not illustrated in Fig.~\ref{fig:03}(b), leading to eigenvalues that are equivalent in the shaded areas by time-reversal symmetry.
This mechanism ultimately enables to open a radiative channel for Bloch modes of the dimerized structure associated with the spatial frequency $M'$, a property that cannot be found in the full-disk lattice, and which is the origin of the symmetry-protected Q-BICs~\cite{Yu19,Overvig20}.

The band diagram for the disk lattice of period $P'$ is shown in Fig.~\ref{fig:03}(b).
Numerical simulations are again performed by employing the COMSOL Multiphysics software following the procedure used in some of our previous studies~\cite{Ghahremani23,Ghahremani24} and elsewhere~\cite{Han19,Han21,Chai21,You23}.
Bloch modes are classified according to their polarization characteristics. 
For instance, the TE$_1^x$ mode exhibits strong $H_z$ and $E_x$ components but negligible $E_y$, $E_z$, $H_x$, and $H_y$ components, and the TM$_1^y$ mode has strong $E_z$ and $H_y$ components and negligible $H_x$, $H_z$, $E_x$, and $E_y$ components~\cite{Vaity22}.
Since $R_i = 0$~nm, modes typified by the Bloch wavevector $\vec{k}$ set at $M'$ in the Fourier plane thus  exhibit an exponential decay at increasing values of the spatial coordinate $|z|$, provided its modulus $|\vec{k}|$ were higher than the wavenumber of the surrounding medium.
In other words, these Bloch modes are perfectly localized below the light line. 
Somehow, they become trapped modes that are recently coined as BICs.
In our design, we find two modes of this kind, which are coined as BIC I and BIC II in Fig.~\ref{fig:03} (b);
in fact they are set at wavelengths $\lambda = 1.19\ \mu$m and $\lambda = 1.26\ \mu$m, respectively.
Note that BIC-I is in fact a degenerate TM$_1^x$-TM$_1^y$ mode, whereas BIC-II is a degenerate TE$_1^x$-TE$_1^y$ mode. 
From a fundamental point of view, this is a direct consequency that the symmetry group C$_{4v}$ is associated with our photonic structure~\cite{Yu19,Overvig20}.
The spatial distribution of the electromagnetic fields for the above-mentioned Bloch modes at the $x$-$y$ midplane of the cylindrical resonators are depicted in Fig.~\ref{fig:03}(c) and (d). 
The orthogonality of each degenerate modal doublet under investigation is apparent from their remarkably similar field distributions, which are rotated by 90 degrees relative to one another. 
The field flow (represented by arrows) indicates that the electric field magnitude is equal in neighboring atoms of the unit cell, but oriented in opposite directions. 
Consequently, the induced currents are fully compensated and cannot couple to the continuum.
In fact, this is the reason these type of dark modes are also called as symmetry-protected BICs.
Let us conclude that the latter argument can alternatively be supported by a vanishing coupling strength of the two BICs with a given normally-incident plane wave by verifying $\gamma_e = 0$, where the coupling coefficient~\cite{Overvig20,Algorri22}
\begin{equation} \label{eq:02}
	\gamma_e
	\propto
	\iint_A
	(\vec{E}_\text{inc}^* \times \vec{H}_\text{mode} + 
	\vec{E}_\text{mode} \times \vec{H}_\text{inc}^*)
	\cdot
	\hat{z}
	\text{d}x \text{d}y ,
\end{equation}
where $A$ is the area of a unit cell.
Note that subscripts ``inc'' and ``mode'' set in the electric and magnetic complex fields refer to the incident and trapped mode, respectively, and $\hat{z}$ is the unit vector in the plane perpendicular to the metasurface.

Next we analyze the band diagram for the holey-disk nanostructure of lattice constant $P$ that is shown in Fig.~\ref{fig:04}(a).
Considering a square array of Si disks with $R_i = 40$~nm, modes with spatial frequencies at $M'$ are then shifted at the $\Gamma$ point within the FBZ, potentially opening a leakage channel through far field radiation.
For TM-like and TE-like polarization, we now find a couple of quasi BICs, coined as Q-BIC I and Q-BIC II, respectively, which are found at $\lambda = 1.17\ \mu$m and $\lambda = 1.25\ \mu$m.
Again, in virtue of the retained C$_{4v}$ symmetry of the photonic structure, both Q-BICs are degenerate modes.
In these cases, however, the eigenfrequencies have a non-vanishing imaginary part leading to a finite $Q$ factor. 
The spatial distribution of the electromagnetic fields for the above-mentioned Bloch modes are depicted in Fig.~\ref{fig:04}(b) and (c).
It is straighforward to recognize that the field patterns are similar to those shown in Fig.~\ref{fig:03}(c) and (d).
On the other hand, the coupling coefficient $\gamma_e$ to our Q-BICs, as defined in Eq.~\eqref{eq:02}, will no longer vanish. 

In order to get further insights about the radiation signature of the Q-BICs, we provide an analysis that is derived from the far-field multipolar decomposition and near-field electromagnetic distribution of an isolated super-cell, which enables uncovering the main contributions of different multipoles on Q-BIC I and Q-BIC II resonances.
The mathematical treatment here accepted for the evaluation of the scatterered power under the multipolar analysis is given as follows:
we computed the scattering cross-section spectra of the Cartesian multipoles in free space by means of the analytical approach~\cite{Tian16,Terekhov17,Zenin17,Babicheva21}:
\begin{align}
	\sigma_\mathrm{sca}
	\approx&
	\frac{k_0^4}{6 \pi \epsilon_0^2 |\vec{E}_\mathrm{inc}|^2}
	\left|
	\vec{P} + \frac{i k_0}{c} \vec{T}
	\right|^2
	+
	\frac{k_0^4 \mu_0}{6 \pi \epsilon_0 |\vec{E}_\mathrm{inc}|^2}
	\left|
	\vec{M} 
	\right|^2
	\nonumber
	\\
	&+
	\frac{k_0^6}{720 \pi \epsilon_0^2 |\vec{E}_\mathrm{inc}|^2}
	\sum_{\alpha \beta}
	\left|
	Q^{(e)}_{\alpha \beta}
	\right|^2
	\nonumber
	\\
	&+
	\frac{k_0^6 \mu_0}{80 \pi \epsilon_0 |\vec{E}_\mathrm{inc}|^2}
	\sum_{\alpha \beta}
	\left|
	Q^{(m)}_{\alpha \beta}
	\right|^2 ,
\end{align}
where $\epsilon_0$, $\mu_0$ and $c$ refers to electric permittivity, magnetic permeability and speed of light in free space, respectively, $\vec{E}_\mathrm{inc}$ stands for the incident electric field and $\alpha, \beta = x, y, z$.
The terms 
\begin{subequations}
\begin{align}
	\vec{P}
	&=
	\frac{i}{\omega}
	\int_\Omega \vec{j} \text{d}v ,
	\\
	\vec{M}
	&=
	\frac{1}{2}
	\int_\Omega \vec{r} \times \vec{j} \text{d}v ,
	\\
	\vec{T}
	&=
	- \frac{1}{10}
	\int_\Omega 
	\left[
	2 r^2 \vec{j}
	-
	(\vec{r} \cdot \vec{j}) \vec{r}
	\right]
	\text{d}v ,
	\\
	Q_{\alpha \beta}^{(e)}
	&=
	\frac{3 i}{\omega} \int_\Omega 
	\left[ 
	{r}_\alpha {j}_\beta + {j}_\alpha {r}_\beta 
	-
	\frac{2}{3} (\vec{r} \cdot \vec{j}) \delta_{\alpha \beta} 
	\right]
	\text{d}v ,
	\\
	Q_{\alpha \beta}^{(m)}
	&=
	\frac{1}{3} \int_\Omega 
	\left[ 
	(\vec{r} \times \vec{j})_\alpha {r}_\beta 
	+ 
	{r}_\alpha (\vec{r} \times \vec{j})_\beta ,
	\right]
	\text{d}v ,
\end{align}
\end{subequations}
are the moments of electric dipole (ED), magnetic dipole (MD), toroidal dipole (TD), electric quadrupole (EQ), and magnetic quadrupole (MQ), respectively, which are evaluated within the scatterer body $\Omega$ through the induced displacement-related current density,
$\vec{j} (\vec{r})
= 
- i \omega \epsilon_0 
\left[
\epsilon (\vec{r}) - 1
\right] 
\vec{E} (\vec{r})$,
which is proportional to the electric field.
The complex-vector electric field within the super-cell is directlly taken from the same numerical simulations used for the evaluation of transmittance, which were previously discussed in Fig.~\ref{fig:01}.

We apply the multipolar analysis to the scattered power by a diatomic unit-cell with $R_i = 40$~nm provided the metasurface is illuminated by a $y$-polarized plane wave.
At the peak reflectance of Q-BIC I, occurring at a wavelength of $1179$~nm, the toroidal dipole components are dominant and, by a lesser significance, the magnetic dipole components, as shown in Fig.~\ref{fig:05}(a).
To gain a more comprehensive understanding of the contributions made by the TD response, we perform calculations to decompose the TD components along different directions, shown in Fig.~\ref{fig:05}(b).
This analysis reveals that that TD$_x$ is governing the far-field scattering of the dimerized ``molecule'' constituting the super-cell.
The near-field distributions of the resonance at the central cross-sections of the $y$-$z$ and $x$-$y$ planes, as depicted in Fig.~\ref{fig:05}(c) and 5(d), respectively, provide confirmation of the observed behavior. 
In the $y$-$z$ plane, the formation of a closed loop by the electric dipoles, predominantly dominated by the $z$ components, signifies the generation of a transverse electric TD resonance oscillating along the $x$-direction~\cite{Jeong20,Wang20}. 
Furthermore, the circular configuration of the electric dipoles surrounding the center of the nanodisks, as shown in Fig.~\ref{fig:05}(d), demonstrates the excitation of a MD oscillating in the same direction as the electric TD.
As expected, it is evident the high degree of similarity between the magnetic field distribution at resonance with the TM$_1^x$ eigenmode of the metasurface given in Fig.~\ref{fig:04}(b).
On the contrary, the behaviour of Q-BIC II is different.
At the resonant wavelength $1249$~nm, the dominant contribution to the scattered field comes from the magnetic quadrupolar term, as shown in Fig.~\ref{fig:05}(e).
Furthermore, the term MQ$_{yz}$ is dominant, as revealed in Fig.~\ref{fig:05}(f). 
Figure~\ref{fig:05}(g) displays the magnetic field distribution in the $x$-$z$ plane, primarily influenced by the $z$-component, providing strong evidence for the excitation of a dominant magnetic quadrupole mode. 
The closed loops observed in the figure indicate displacement currents oriented perpendicular to the plane, characteristic of the quadrupolar nature of the magnetic field. 
Figure~\ref{fig:05}(h) complements this discussion by illustrating the distribution of these displacement currents in the $x$-$y$ plane.
Finally, the substantial correlation between the electric field distribution of the scattered near-field shown in Fig.~\ref{fig:05}(h) and that of the TE$_1^y$ eigenmode of the metasurface given in Fig.~\ref{fig:04}(c) is perceptible.

The analysis given above is based on a specific direction of polarization of the impinging plane wave. 
While not shown graphically, the differences found between bringing into action an incident optical signal with $y$ polarization, as discussed above, and $x$ polarization come down to a rotation of the electromagnetic fields by 90 degrees.
This is a characteristic feature of metasurfaces with geometries that belong to the symmetry group C$_{4v}$~\cite{Yu19,Overvig20}, revealing the modal degeneracy of the Q-BICs excited in this sort of photonic structures. 
Obviously, the Cartesian components of the multipole expansion undergo a cyclic permutation under polarization rotation, but the overall contribution is maintained fixed.
This fact further supports the polarization-independent behaviour of the metasurface.
Figure~\ref{fig:06}(a) shows the transmission spectra by changing the polarization angle $\varphi$ with respect to the $x$ axis at $R_i = 40$~nm, evidencing an invariant response.
We conclude by expanding our discussion about the polarization independence of Q-BIC I and Q-BIC II.
Figure~\ref{fig:06}(a) also depicts the in-plane distributions of magnetic and electric fields for both TM-like and TE-like resonant modes at an incident angle $\varphi = 45^\circ$. 
In Fig.~\ref{fig:06}(b) and (c), the transmission spectrum and field profiles of the metasurface are illustrated for right-handed (RH) and left-handed (LH) circularly polarized (CP) light, respectively. 
Notably, one can realize that the transmission spectrum exhibits identical profiles across all polarization states. 
Furthermore, the observed electric and magnetic near-fields align with those of the BIC eigenmodes, definitively characterizing the resonance.

\section{Concluding remarks}

We discussed a novel ``dimerized'', highly-symmetric, dielectric metasurface with dual-mode resonances governed by symmetry-protected BICs. 
Particularly, we analyze the effect of the Brillouin zone folding on the existence of BICs and reveal polarization-independent response of the quasi-BIC states in addition exhibiting extraordinary $Q$ factors, which are supported by the resulting “diatomic” dielectric metasurfaces.
This metasurface is insensitive to normally incident light with any polarization state and offers higher $Q$ factors than other similar proposals. 
Our procedure involves perturbing the metasurface structure by introducing a concentric hole in half of the nanodisks, maintaining the C$_{4v}$ symmetry group of the structure and leading to the formation of Q-BIC states. 
The numerical simulations showcase the metasurface transmittance under different conditions. 
This study references related works on all-dielectric near-infrared photonics and extended Q-BICs with non-reduced symmetry in NIR dielectric metasurfaces.
The efficient design provided by the method here described is competitive with that of most of the existing methods and are sufficient for most applications including nonlinear optics and structural colors, to name a few.
We particularly suggest the potential for ultrahigh-performance sensing applications with simpler fabrication processes.


\textbf{Disclosures.} 
The authors declare no conflicts of interest.

\textbf{Data availability.} 
Data underlying the results presented in this paper are not publicly available at this time but may be obtained from the authors upon reasonable request.




%

\end{document}